\documentclass{WileyMSP-template}
\usepackage{soul}
\usepackage{xcolor}
\usepackage{mathtools}

\begin{document}

\pagestyle{fancy}
\rhead{\includegraphics[width=2.5cm]{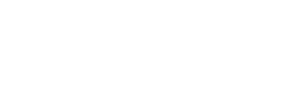}}

\title{Dynamics of Vector Soliton Singlets and Pairs in Self-Mode-Locked Tm-Doped Fibre Lasers}

\maketitle

\author{Dennis C. Kirsch*}
\author{Anastasia Bednyakova}
\author{Maria Chernysheva}

\dedication{}

\begin{affiliations}
D. C. Kirsch\\
Leibniz Institute of Photonic Technology, Albert-Einstein-Straße 9, 07745, Jena, Germany\\
Email Address: dennis.kirsch@leibniz-ipht.de

Dr. A. Bednyakova\\
Novosibirsk State University, Novosibirsk, 630090, Russia\\
Email Address: anastasia.bednyakova@gmail.com

Dr. M. Chernysheva\\
Leibniz Institute of Photonic Technology, Albert-Einstein-Straße 9, 07745, Jena, Germany\\
Email Address: Maria.Chernysheva@leibniz-ipht.de

\end{affiliations}


\begin{abstract}
Vector solitons present unique optical pulses characterised by coupled orthogonal polarisation components, offering rich, multidimensional dynamics. Their enhanced degrees of freedom make them highly versatile for exploring nonlinear wave phenomena and enabling advanced applications in optical communications and ultrafast technologies. Despite significant recent progress, the intricate dynamics of vector soliton generation remain largely unexplored, particularly their interaction and coupling mechanisms. This study characterises soliton build-up dynamics in a self-mode-locked Tm-doped fibre laser, where both cavity birefringence and distributed saturable absorber play significant roles in the intracavity pulse shaping. The recorded single-shot spectra using the Dispersive Fourier Transform technique reveal generation regimes of singlet solitons and loosely bound two-pulse states. In later case, the observed intensity-dependent temporal splitting of vector soliton polarisations is also affected by the repulsive forces due to the interaction with synchronised and unsynchronised dispersive waves, governing the transition from transient pulse formation to steady-state generation. This study enhances our understanding of vector soliton dynamics and sets a solid foundation for further advanced exploration of their generation regimes across a wide range of nonlinear photonic systems.

\end{abstract}


\section{Introduction}

The field of ultrafast photonics has been rapidly developing over the last decades, driven by their wide range of applications~\cite{fermann2002ultrafast, phillips2015ultrafast, braun2008ultrashort} and fundamental studies. Thus, ultrafast lasers present a universal platform for investigations of nonlinear wave dynamics, which includes the observation of solitons~\cite{herink2016resolving,blanco2023bright} or rogue waves formation~\cite{onorato2013rogue}, the interaction of coherent structures with noise~\cite{copie2020physics}, to name a few. Since the first introduction of the concept of overlapping soliton interactions under pumping-dissipation conditions~\cite{malomed2005bound}, increasing attention has been devoted to the coupling mechanisms of multiple solitons. Thesoliton-soliton interactions can experience different attraction and repulsive forces, resulting in diverse configurations~\cite{liu2018real,krupa2017real}, including their various stationary variations, such as macro-molecules, soliton crystals or supramolecular structures~\cite{he2019formation}, and breathing multi-pulse structures~\cite{wang2019real,peng2021breather}. Soliton interactions find their analogy with matter-wave interactions, which can reach equilibrium when the opposing forces of attraction and repulsion are perfectly balanced. Furthermore, bound solitons offer enhanced degrees of freedom for emerging applications, particularly for coding information storage applying machine learning algorithms~\cite{si2024deep,luo2024real}.

Vector soliton pairs represent another distinct robust form of multi-pulse state. Dual polarisation components inherent to vector solitons present additional degrees of freedom to control ultrashort pulse properties, enabling a range of applications in ultrafast switching, signal multiplexing, and polarisation-sensitive sensing. The manifold of vector soliton dynamics patterns has been primarily studied numerically~\cite{menyuk1987nonlinear,menyuk1987stabilityI}. At the same time, only the cases of trapped group velocity-locked~\cite{islam1989soliton} and the phase-locked vector solitons~\cite{cundiff1997polarization} typically with the equal strengths of the partial pulses in each of the two polarisations have been reported experimentally. In fact, by controlling the strength of orthogonal polarisation and cavity birefringence, more complex pulse interactions can be triggered, yielding, for example, the formation of breathing locked pairs of vector solitons~\cite{menyuk1988stabilityII}. While the experimental observation of the generation of such pulse complexes and their nonstationary evolution presents an unfeasible task for the averaged measurements, the real-time measurement methodologies, such as time-stretch Dispersive Fourier Transform spectroscopy (DFT)~\cite{godin2022recent,goda2013dispersive}, brings the possibility of single-shot observation of fast non-repetitive dynamics at time scales of the laser cavity roundtrip~\cite{du2019pulsating,krupa2017vector,zhao2021real}. 

The advantageous concept of the DFT measurements allows single-shot recording of an optical spectrum of the pulse, exploiting the space-time duality analogously to the paraxial diffraction~\cite{jannson1983real}. 
The most straightforward and broadband method to implement the dispersive pulse broadening, fulfilling the far-field conditions, is by applying a spool of optical fibre~\cite{jannson1983real}. Due to the availability of low-loss optical fibres, this method has been widely used to study dynamics in Er-doped fibre lasers~\cite{goda2013dispersive,mahjoubfar2017time,kudelin2020pulse,peng2018real,liu2018real}. The implementation of DFT measurements at other wavelength ranges, e.g., beyond 1.7~$\mu$m, remains relatively rare due to the phonon absorption and requires more elaborated pulse stretching techniques. Thus, an application of a set of chirped fibre Bragg gratings~\cite{hamdi2018real,zhou2022reconfigurable}, a recirculating photonic filter using arrayed waveguide gratings with an arbitrary delay at each spectral channel, multi-mode fibres and waveguides ensuring chromomodal dispersive broadening has been reported over recent years~\cite{mahjoubfar2017time}. At the same time, the absence of fast photodetectors at longer wavelength ranges~\cite{filatova2024experimental} has been mitigated through the second harmonic generation in nonlinear crystals~\cite{zeng2022real,huang2019route}.

Nevertheless, ultrafast laser operation at longer wavelengths presents numerous opportunities not only for a wide range of applications~\cite{kirsch2020short,wang2020recent} but also for triggering exotic pulse dynamics. In this work, we study the formation evolution of vector solitons in a Tm-doped ring fibre laser in real-time in the temporal and spectral domains. The ultrafast fibre laser employs nonlinear energy exchanges in Tm-ion pairs, forming a distributed saturable absorber. First, we demonstrate the build-up dynamics of a conservative-like soliton. In the case of higher pump power, due to the high intracavity birefringence accompanied by the emerging dispersive waves, the pulse polarisation components split in the temporal domain and experience repulsive interactions, yielding the formation of dual soliton structures. The experimental observations are verified by numerical simulations of the presented laser systems using coupled modified nonlinear Schr\"{o}dinger equations. The results deepen the understanding of nonlinear dynamics in ultrafast fibre lasers mode-locked through Tm ion–pair interaction and vector solitons interactions through the dispersive waves.

\section{Results}

The self-mode-locked Tm-doped fibre laser, presented in detail in the Methods section~\cite{kirsch2022gain}, generates a stable train of soliton pulses with the pump power threshold of $\sim$330~mW. After a stable self-starting mode-locking is achieved in the laser system by adjusting the polarisation controllers, the pump is turned off and on again, with an oscilloscope triggered to record the generated signal. Notably, the laser cavity comprises standard isotropic optical fibres and components, i.e. it does not maintain a particular polarisation state. With the polarisation state as one of the degrees of freedom for the vector solitons dynamics, the self-mode-locked laser shows a rich assortment of pulse regimes. The pulse build-up dynamics were recorded in different laser settings of pump power (namely, at 400 and 470~mW) or polarisation controller position, meaning overall cavity birefringence. It is essential to notice that the variation in pump power or oscillating energy in the cavity alters the ratio of pumped and unpumped sections of gain fibre, thus affecting the saturable absorber parameters~\cite{kirsch2022gain} and, as a result, the transient dynamics.




\subsection{Singlet soliton formation}


\begin{figure*}[b!]
\centering
\includegraphics[width=.98\linewidth]{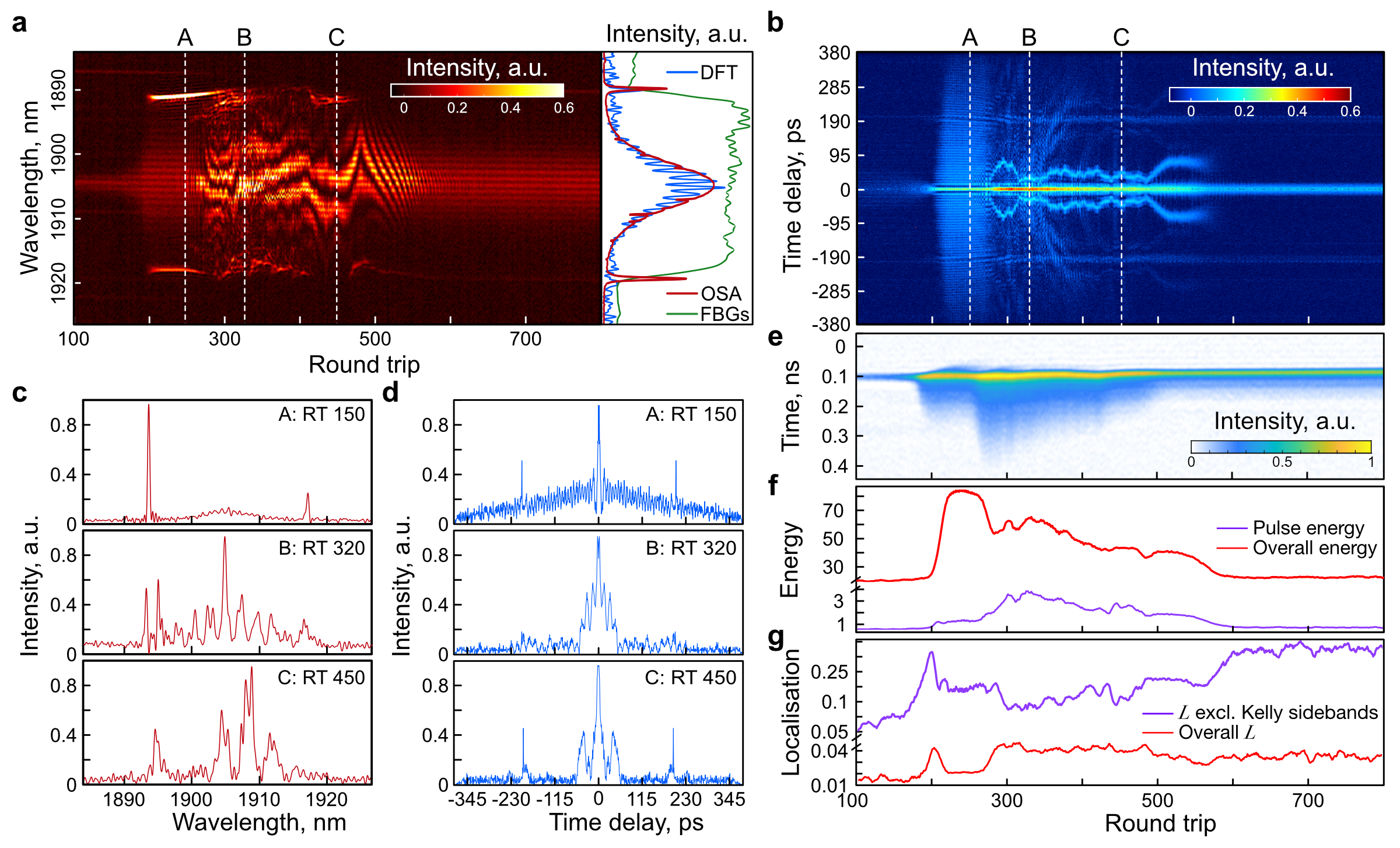}
\caption{\textbf{Experimentally recorded formation dynamics of fundamental soliton in self-mode-locked Tm-doped fibre laser. a} Single-shot spectral evolution with characteristic cross-sections in \textbf{c}. Side panel: Averaged over 10 round trips DFT spectra in the steady-state, spectrum recorded via optical spectrum analyser (OSA) and transmission spectrum of the used CFBGs for dispersive stretching. \textbf{b} Field autocorrelation evolution with corresponding cross-sections in \textbf{d}. \textbf{e} Spatio-temporal evolution; \textbf{f} Dynamics of the round trip pulse energy and \textbf{g} localisation parameter. }
\label{fig:solitonExp}
\end{figure*}

\textbf{Figure~\ref{fig:solitonExp}}(a) provides a complete overview of pulse formation dynamics recorded using the DFT technique at pump power slightly above the mode-locking threshold of 400~mW. The DFT measurement setup is described in the Methods section. The single-shot spectra evolution depicted in Figure~\ref{fig:solitonExp}(a,b) reveals four distinct phases of the formation dynamics. After approx.~500-600 round trips, the laser reaches the steady-state operation with an average power of around 01mW and output optical spectra, depicted as a red plot on the side panel of Figure~\ref{fig:solitonExp}. Here, the modulation of single-shot pulse spectra (blue plot), recorded via DFT, is attributed to the reflection spectrum modulation of the chirped fibre Bragg gratings (CFBGs) used for dispersive stretching of the pulse (green plot). The field autocorrelation trace, depicted in Figure~\ref{fig:solitonExp}(b), also confirms the absence of the side pulse and single pulse generation in the steady-state. The field autocorrelation trace has been obtained using the Wiener–Khinchin theorem by performing a fast Fourier transform of each single-shot spectrum. Figures~\ref{fig:solitonExp}(c,d) demonstrate characteristic single-shot spectra and field autocorrelation traces corresponding to the pulse formation stages. 

Similar to formation dynamics shown previously~\cite{herink2016resolving,peng2018real,kudelin2020pulse}, the pulse build-up starts through pronounced modulation instability, featuring an ensemble of stochastically distributed picosecond-lasting, single wavelength, quasi-cw noise spikes. Via discretionary superposition between several spikes, the dynamics evolve to wavelength broadening and the formation of a transient pulse with a high ratio of energy spread broadly over a dispersive wave background, as shown in cross-sections at RT~150 (cross-section A). Figures~\ref{fig:solitonExp}(f,g) demonstrate dynamics of the pulse energy and its distribution across the round trip time span, correspondingly. According to Parseval's theorem, the energy fluctuation has been estimated by integrating the single-shot spectra from the DFT measurements. We used the localisation parameter introduced in~\cite{kudelin2020pulse} to evaluate energy distribution as a ratio of integrated field autocorrelation trace within the pulse intensity profile to overall energy. The energy of the dispersive wave exceeds that of the transient pulse by more than an order of magnitude. High background radiation can also be seen in the field autocorrelation trace and spatio-temporal dynamics in Figure~\ref{fig:solitonExp}(e). As a result, the emission of a dispersive wave destabilises the pulse, resulting in its splitting after nearly 280 roundtrips. 

The interference pattern of the spectral dynamics in Figure~\ref{fig:solitonExp}(a) manifests the formation of bound solitons (as shown in cross-sections B and C). The field autocorrelation trace in Figure~\ref{fig:solitonExp}(b) reveals the fission of the transient pulse into two main solitons at this position. Since the secondary solitons do not feature balanced attraction to the main pulse, they move temporally away from the central soliton, periodically recur, and get ricocheted several times from the central soliton. The secondary soliton extinguishes after about 600 round trips (cf.~ref.~\cite{roy2005dynamics}). This clearly illustrates the action of the intensity-saturating attenuation in the oscillator, which, together with the gain, permits only the viability of the stronger soliton after a certain amount of round trips. Figure~\ref{fig:solitonExp}\,(f) indicates that between 200 and 600 roundtrips, most energy is still spread in the dispersive background wave. This can also be seen in the spatio-temporal dynamics of the pulse, presented in Figure~\ref{fig:solitonExp}(e). Interestingly, both overall energy and the energy localised in the central pulse decrease towards the end of the bound soliton stage of formation dynamics. 

A stable equilibrium emerges at approx. 600 round trips, resulting in the final stable pulse propagation. In the right-side panel of Figure~\ref{fig:solitonExp}\,(a), the final stable pulse is shown as the average of ten round trips (blue plot). It spans over a full-width half-maximum of ~nm. Due to the limited bandwidth of the used CFBGs for dispersive stretching, the long-wavelength Kelly sideband is highly attenuated and almost indistinguishable in the single-shot spectra recorded via DFT. 
The steady-state features a clean field autocorrelation without secondary peaks and exhibits a deconvoluted pulse duration of approximately 6~ps (Figure~\ref{fig:solitonExp}\,b). As shown in Figure~\ref{fig:solitonExp}(g), with the annihilation of secondary pulses, the localisation parameter, estimated with excluded Kelly sidebands, shows a steady uptrend, reaching the proximity of 0.4. This value is the theoretical limit for the case when all energy is localised within ${sech^2}$-shaped pulse. Hence, such dynamics justify that the final pulse is a well-developed fundamental soliton.



\begin{figure*}[!bp]
\centering
\includegraphics[width=.98\linewidth]{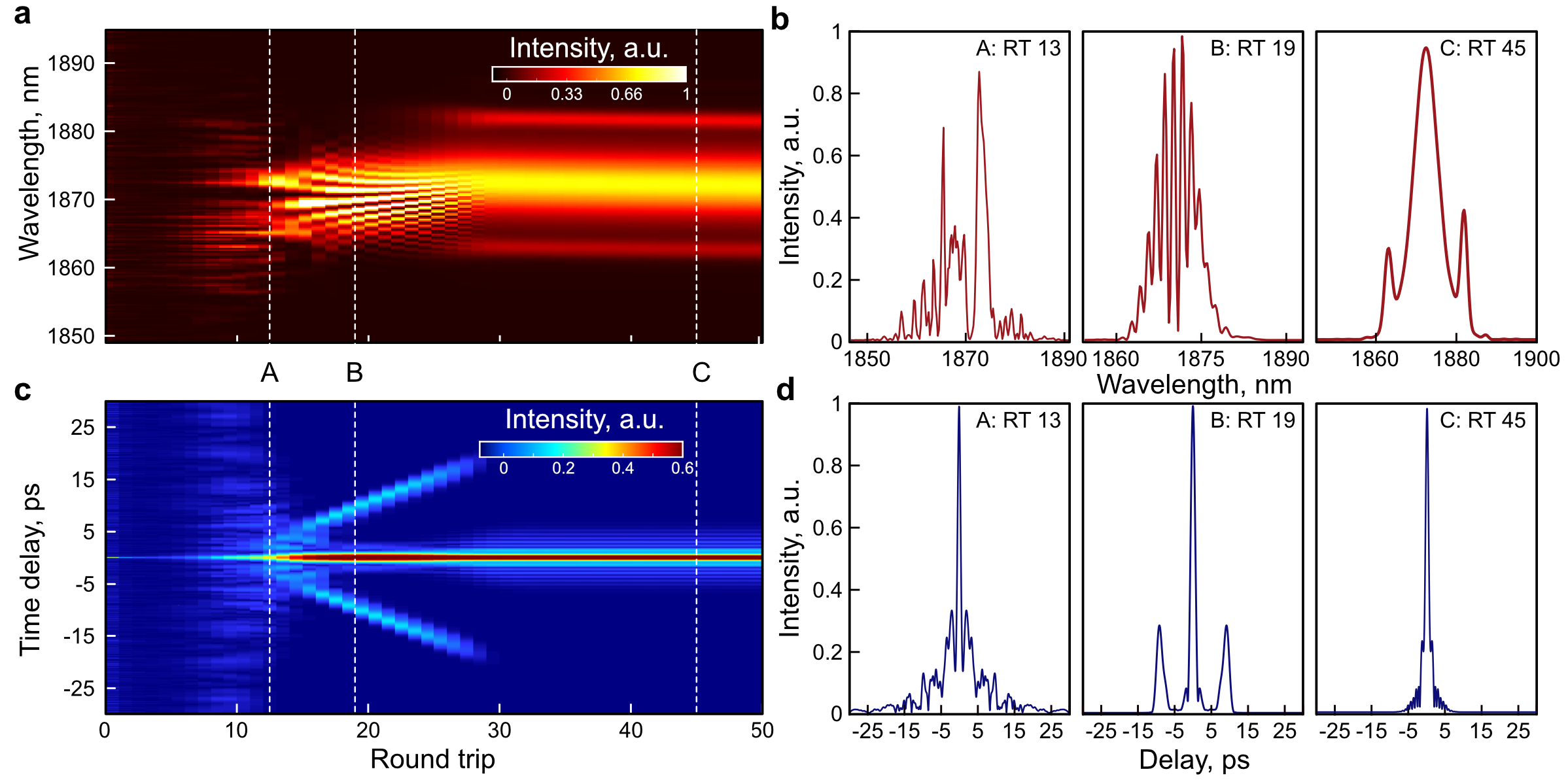}
\caption{\textbf{Numerical simulation of formation dynamics of fundamental soliton in self-mode-locked Tm-doped fibre laser. a} Spectral dynamics with corresponding cross-sections in \textbf{b}; \textbf{c} Field autocorrelation evolution with characteristic cross-sections in \textbf{d}.}
\label{fig:solitonTheor}
\end{figure*}

Solving the nonlinear Schrödinger equation numerically, as described in the Methods section, using the laser parameters from~\cite{kirsch2022gain} has revealed the formation dynamics, which qualitatively account for the experimentally observed evolution. \textbf{Figure~\ref{fig:solitonTheor}} demonstrates single-shot spectral and field autocorrelation evolution with characteristic cross-sections, featuring distinct formation stages from MI spikes, followed by intensive dispersive wave formation, through soliton split up to final steady state generation. 



\subsection{Vector soliton pair formation}
\begin{figure*}[!]
\centering
\includegraphics[width=.98\linewidth]{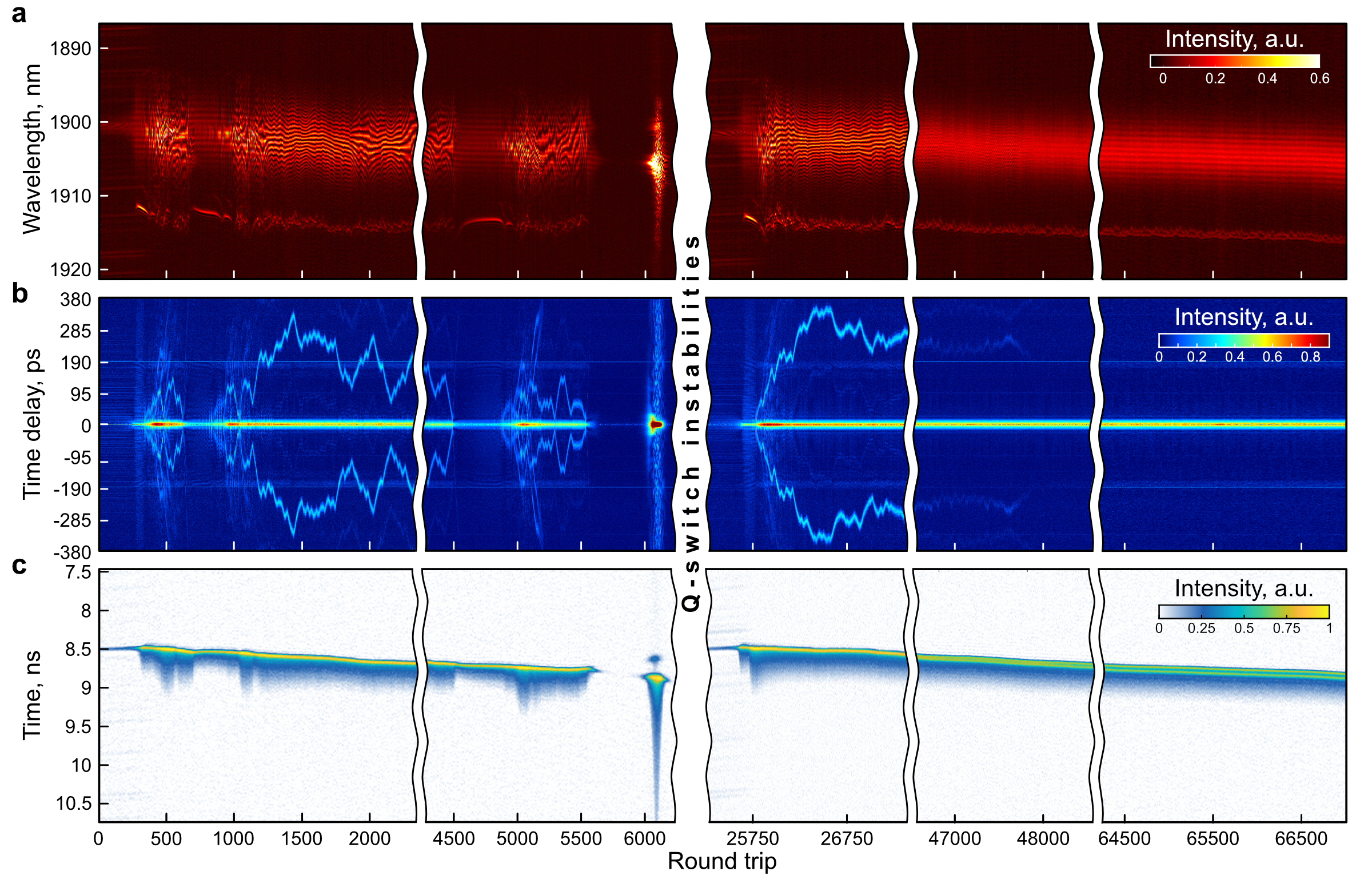}
\caption{\textbf{Build-up dynamics of vector soliton pair: a} Single-shot spectral evolution; \textbf{b} Field autocorrelation evolution; \textbf{c} Spatio-temporal evolution.}
\label{fig:Dual_all}
\end{figure*}

The build-up dynamics become more rich and complex with the increase of pump power to 471~mW within an otherwise unmodified layout. \textbf{Figure~\ref{fig:Dual_all}} provides a closer look at the single-shot spectra (a), field autocorrelation trace (b) and spatio-temporal (c) dynamics. Overall, the initiation of the formation process resembles the above-discussed case of the single soliton build-up, starting from modulation instability followed by the emergence of transient pulse with intensive dispersion wave and bound solitons dynamics~\cite{liu2018real}. The formation of quasi-stable solitons is accompanied by a slight wavelength drift (a few nm), which is most pronounced for the long-wavelength Kelly sideband. Notably, the final formation is delayed by several stages of splitting the fundamental pulse into soliton complexes and their recombination and by an extended stage of Q-switched instabilities. Particularly, the Q-switched instabilities occurred after the merging of bound state solitons into an unstable soliton generation at around 5700 round trip. This stage lasted for nearly 20000 round trips (corresponding to 0.4~ms), featuring a Q-switched mode-locking generation regime at an 83-kHz repetition rate. After the Q-switched pulse, the modulation instability again gives rise to a new intensity spike in the dominant direction, and the formation dynamics repeat the previously discussed stages until stable steady-state generation is achieved.

\begin{figure}[!]
\centering
\includegraphics[width=.98\linewidth]{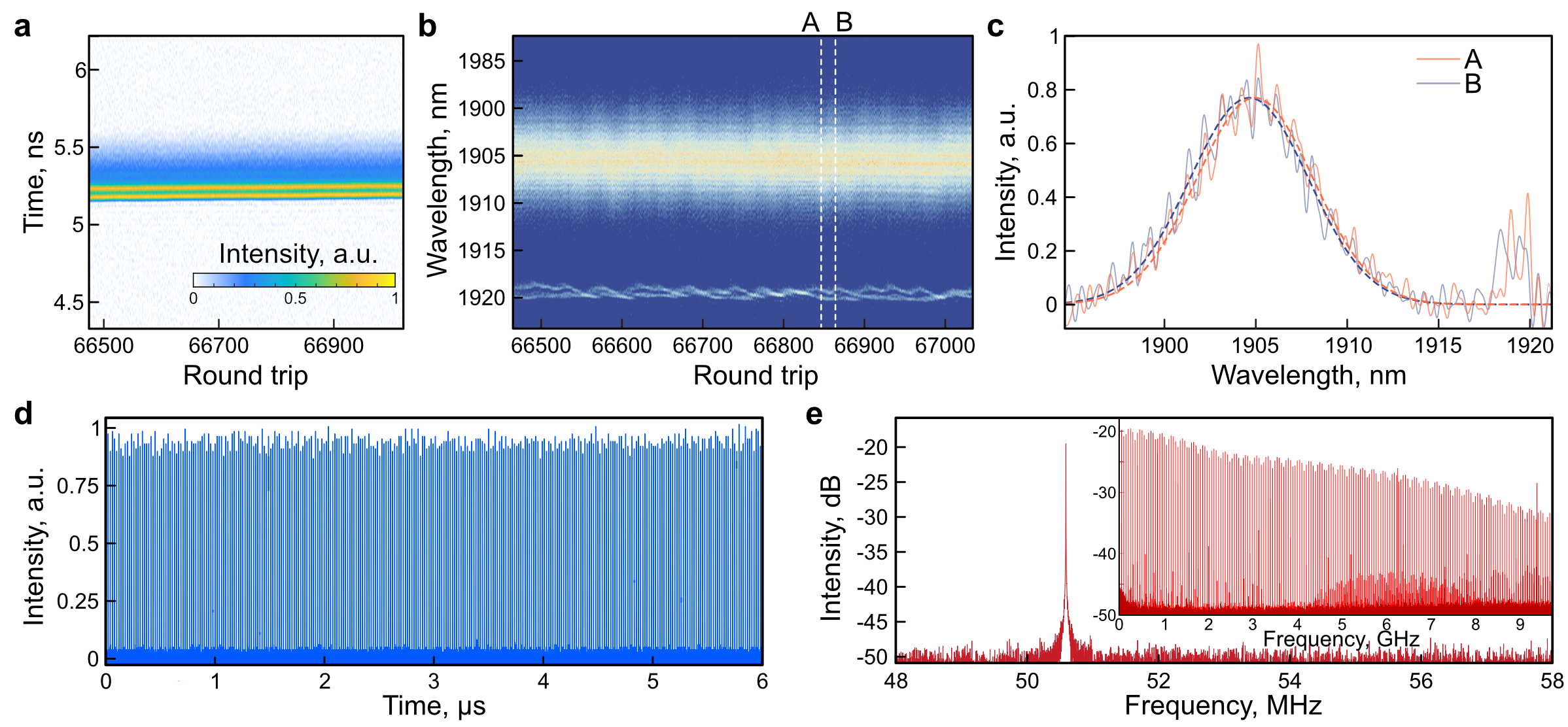}
\caption{\textbf{Steady-state generation of trapped vector solitons in self-mode-locked Tm-doped fibre laser. a} Spatio-temporal evolution; \textbf{b} Dynamics of breathing single-shot spectra with characteristic cross-sections at A and B in \textbf{b}; \textbf{d} Oscilloscope trace, showing $\sim$5\% intensity deviation. \textbf{e} RF spectrum at fundamental pulse repetition frequency. Inset: Zoomed-out RF spectrum with 10~GHz span.}
\label{fig:Dual_FinalStage}
\end{figure}

The steady-state generation is analysed in detail in \textbf{Figure~\ref{fig:Dual_FinalStage}}. It is important to highlight that the bound-state multi-pulse complex vanishes by reaching the steady-state operation at $\sim$48000 round tripe same time, the central pulse splits into two stable, closely positioned trapped vector solitons. The spatio-temporal evolution in panel~(a) justifies the presence of two pulses, separated by $\sim$53~ps. In the usual case of a group of adjacent pulses co-moving at a constant small time interval in the cavity, one would expect a spectral interference pattern to arise based on their superposition~\cite{chernysheva2020real,peng2021breather}. Such a temporal separation of bound solitons at 1905~nm central wavelength would correspond to the fringe period of $\sim$0.22~nm in the single-shot spectra, similar to the interference pattern of soliton complexes recorded during the build-up dynamics. However, despite sufficient resolution, neither DFT measurement nor retrieved field autocorrelation trace, demonstrated in Figure~\ref{fig:Dual_all}(a,b), evidence of the presence of bound soliton complex, as the only spectral modulation is introduced by the profile of the CFBG used for DFT measurements. We attribute the absence of interference between the pulses to their different polarisation~\cite{cundiff1997polarization,zhao2008soliton}. Since the laser cavity comprises only standard isotropic optical fibres and components, the combined influences of a random birefringence of optical fibres and strain-induced fibre birefringence in a coiled fibre section and squeezing in-line fibre polarisation controller force the polarisation state to evolve freely.


Vector solitons are generally generated at different group velocities and then trap each other through the shifts of their central frequency via nonlinear Kerr effect~\cite{menyuk1987stabilityI} so that they can propagate over long distances as a single non-dispersive unit. However, at larger birefringence values and unequal amplitude of vector solitons, the dynamics of the pulses get more complex. Thus, as shown theoretically in Ref.~\cite{menyuk1988stabilityII}, the vector soliton could feature primarily one polarisation together with a certain contribution of another. These polarisation states would move apart from the central position and form a breathing two-soliton structure. With the increase of the pulse intensity, the leading pulse will experience the growth of the amplitude, while the amplitude of the trailing pulse will decrease. This explains the generation regime observed in the experiment, revealing the dual-peak structure of the pulses' temporal profile, shown in Figure~\ref{fig:Dual_FinalStage}(a) and the absence of the interference pattern in the single-shot spectra recorded using the DFT technique (Figure~\ref{fig:Dual_FinalStage}(b). 

As seen in Figure~\ref{fig:solitonExp}(e) and ~\ref{fig:Dual_all}(c), the formation dynamics in both cases of lower and higher pump power are affected by the intensive dispersive wave. The initial low pump power case demonstrates clearly that the intensity and width of the dispersive wave decrease with the annihilation of the soliton complex and the promotion of only the principal pulse peak. In the case of the higher pump power, the dispersive wave decreased during the states of the formation of the transient pulse at $\sim$700-900 and 4500-4900~round trips. However, the transient pulse could not be supported in the given laser settings and broke into the soliton complexes associated with the intensive dispersive waves. It is important to notice that the presence of the intensive dispersive wave promotes the repulsive interactions, facilitating the formation of a loosely bound pulse complex with a larger separation between the solitons that are still coupled to each other~\cite{soto2003quantized}, similar to ones the observed in the current work experimentally. 

As seen in Figure~\ref{fig:Dual_FinalStage}(b), the recorded single-shot spectra oscillate along roundtrips, identifying that the dual-pulse complexes restore their original spectrum every 46~roundtrips. The characteristic spectra cross-sections, shown in Figure~\ref{fig:Dual_FinalStage}(c), demonstrate oscillation of the central wavelength, yet almost negligible change in the spectral bandwidth. Contrary to previous demonstrations of vector soliton dynamics~\cite{luo2020stationary}, no pulsations have been observed in the intensity of pulses in the temporal domain, which features nearly 5\% stochastic deviation (Figure~\ref{fig:Dual_FinalStage}d). Neither subpeaks could be observed in the radio frequency spectrum, shown in Figure~\ref{fig:Dual_FinalStage}(e), opposite to previous works~\cite{zhao2021real}. The breathing nature of the vector two-soliton complex leads to the intricate interaction with the dispersive wave~\cite{weill2011spectral}, giving rise to the breathing dynamics of Kelly sidebands, as seen in Figure~\ref{fig:Dual_FinalStage}(b). The Kelly sidebands in the soliton spectrum arise from their resonant interaction in the temporal domain with dispersive waves radiated from the pulse when perturbed by intracavity components and lumped nonlinear losses~\cite{komarov2012dispersive}. At the same time, the evolution of the spectrum over cavity round trips does not show a breathing behaviour, such as a change of the bandwidth or wavelength shift, but insignificant variations at the noise level of the measurement equipment. 
There are two components in the long wavelength Kelly sideband, as shown in Figure~\ref{fig:Dual_FinalStage}. One peak corresponds to a synchronised resonant dispersive wave (at a shorter wavelength), as its fast oscillation matches the oscillation of the wings of the spectrum. The other one refers to an unsynchronised resonant dispersive wave with an almost fixed central wavelength, i.e., less pronounced fluctuation. The fast wavelength oscillation of the Kelly sideband indicates that a resonant and an unsynchronised dispersive wave mediate due to different group-velocity dispersions, resulting in period bifurcation. As the dispersive waves are periodically emitted during hundreds of cavity periods by each of the orthogonally polarised pulses, the emergence of unsynchronised Kelly sidebands can be attributed to cross-phase modulation of the solitons and the orthogonal radiated dispersive waves~\cite{du2019pulsating}. 

\begin{figure}[!bp]
\centering
\includegraphics[width=.98\linewidth]{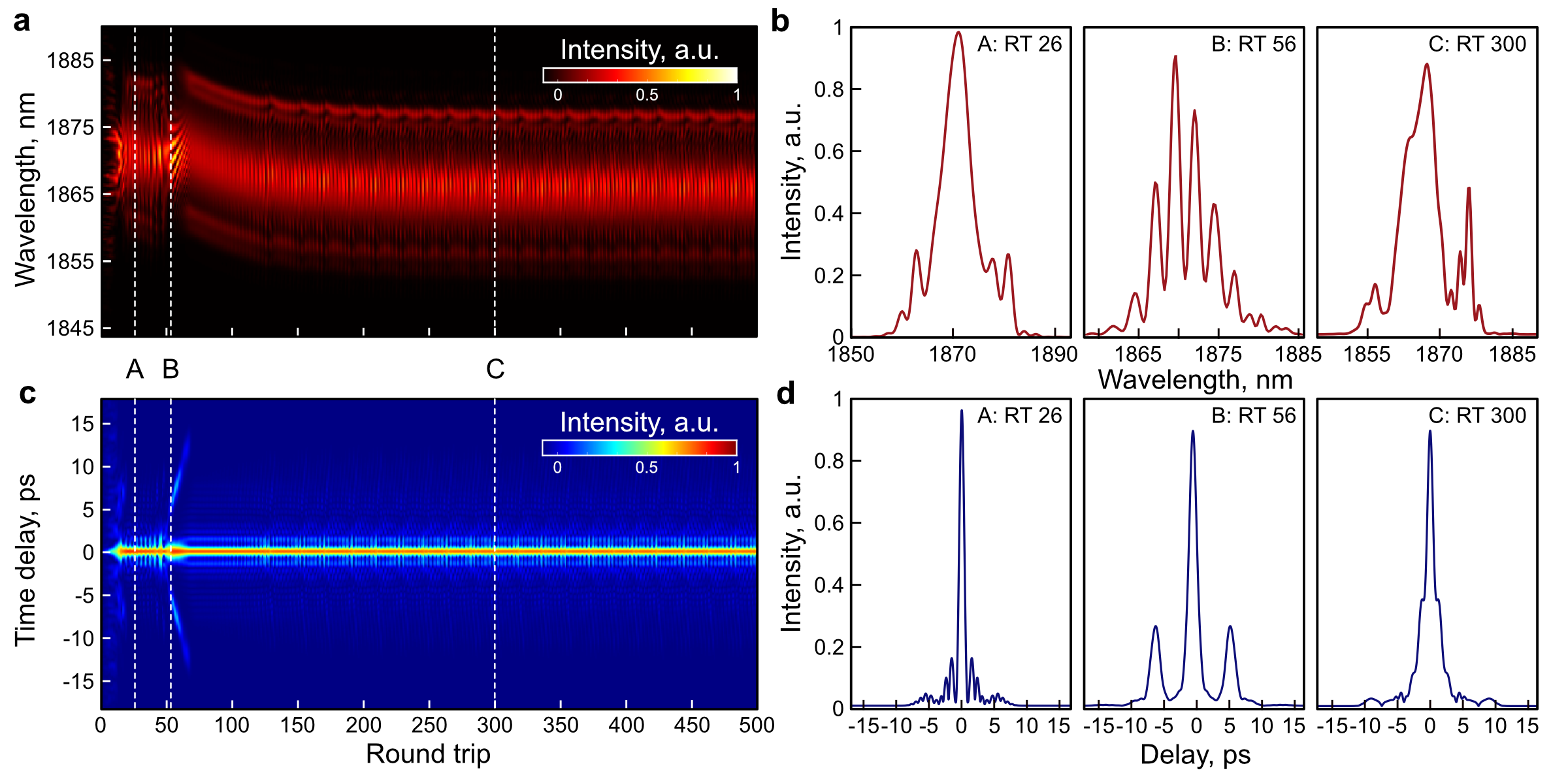}
\caption{\textbf{Numerically simulated vector soliton formation a} Single-shot spectral and \textbf{b} field autocorrelation evolution.}
\label{fig:Theor-vector}
\end{figure}

\textbf{Figure~\ref{fig:Theor-vector}} demonstrates the numerical simulation of formation dynamics of vector solitons, accounting for the effect of cross-phase modulation. The simulation results qualitatively verified the wavelength shift at the beginning of the solitons formation, as well as the formation of dual Kelly sidebands and their oscillating dynamics. It is worth noting that at the early stages of pulse build-up, with the formation of a quasi-stable transient pulse, Kelly sidebands stabilise. However, the same oscillating behaviour remains with the laser transitioning from bound-state soliton operation into the steady-state regime. Therefore, both theoretical and experimental results have confirmed that unless the breathing dynamics of Kelly sidebands are obtained, stable steady-state generation of vector solitons cannot be observed. The pulses otherwise experience break-up into soliton complexes and undergo wavelength shift.

\section{Conclusions}

This work reports the experimental observation and real-time characterisation of soliton build-up dynamics in self-mode-locked Tm-doped fibre lasers. The fibre laser design, comprising all polarisation-independent components and distributed saturable absorber over an unpumped rear section of gain fibre, presents a beneficial platform for the investigation of vector soliton generation, trapping, and weak interactions over pulse build-up and steady-state generation. The developed numerical model of the self-mode-locked Tm-doped fibre laser has accurately mapped the complex cavity parameters. This allows for an excellent qualitative verification of the experimentally observed vector solitons formation dynamics. We demonstrated what we believe to be an earlier unobserved experimental regime of operation associated with intensity-dependent temporal splitting of vector soliton polarisations. Together with the repulsive forces introduced by an intensive dispersive wave at laser pump powers, it promotes splitting a fundamental soliton into multi-pulse emission, forming robust vector two-pulse loosely bound states. Furthermore, our results have revealed the generation of synchronised and unsynchronised dispersion waves and showed their role in establishing the steady-state generation and arising complex pulse spectra behaviour. We anticipate our work to enrich the understanding of the vector dynamics of pulsating solitons in ultrafast fibre lasers and stimulate their further studies in various dissipative systems.

\section{Methods}

\subsection{Experimental setup}
\begin{figure}[!bp]
\centering
\includegraphics[width=0.98\linewidth]{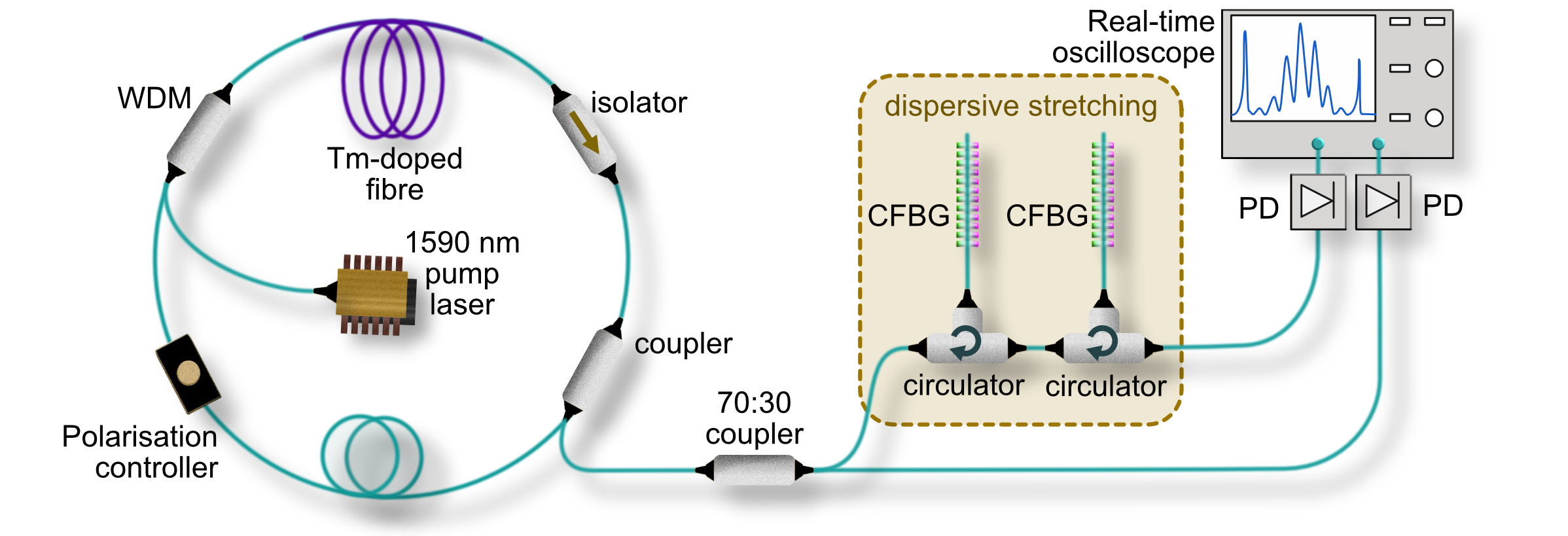}
\caption{Setup of the laser under test and the DFT measurement equipment.}
\label{fig:setup}
\end{figure}

The investigated fibre laser, depicted in \textbf{Figure~\ref{fig:setup}}, comprises an all-anomalous dispersion ring cavity, 
based on a 48-cm gain fibre doped with about 1.8$\cdot$10$^{26}$\,Tm$^{3+}$\,m$^{-3}$~\cite{kirsch2022gain}. Having approximately 20\% of these ions clustered in pairs, such doping concentration has been found to provide a compromise between the high laser gain due to homogeneously distributed Tm\textsuperscript{3+} and effective saturable absorption in ion clusters in the rear un-pumped fibre section. The origin of the advantageous saturable absorption in rare-earth-doped fibres is still yet to be thoroughly investigated. It is associated with ion pairs excited-state absorption \cite{le1993influence,El-Sherif.2002d,tang.2011m} and upconversion interaction  \cite{sanchez1993effects,colin1996evidence}. In the context of Tm-doped fibres, $^3F_4$, $^3F_4$~$\rightarrow$~$^3H_6$, $^3H_4$ energy transfer upconversion plays a key role in establishing the saturable absorption behaviour~\cite{Jackson.1999d}. 

A polarisation-insensitive isolator and an in-line polarisation controller force a clockwise pulse propagation and a consistency of the polarisation state, respectively. The laser cavity comprised only standard isotropic optical fibres and components, introducing a weak linear birefringence. According to a specified statistical lateral refractive index, the variance of used passive fibre (standard G.652) is $\sim10^{-6}\;n$, and slightly larger for used active fibre. Therefore, the polarisation beat length $L_b= \lambda\; \Delta{n}_{\it eff}^{-1}$, the length over which the phase difference between orthogonal polarisation increases by $2\pi$, can be assumed to lay in the order of 1.9~m. Furthermore, to fix the laser generation wavelength within the bandwidth of the CFBGs used for DFT measurements, a section of passive fibre was coiled into two circles with a radius of 35~mm. The stress-induced birefringence due to the coiling of the fibre at $\lambda = 1905 \mu$m is $\beta_B = -2.56 \cdot 10^7 k^2 r^2$~deg/m, where $k = 1/R$ -- is the coiling curvature and $r$ -- is the fibre radius~\cite{ulrich1980bending}. With the fibre cladding diameter of 125~$\mu$m, the bending-induced birefringence is $\beta_B$~=~81.6~deg/m, corresponding to a polarisation beat length of 2.2~m. However, it is quite challenging to estimate the birefringence induced by an in-line squeezing polarisation controller.

For the pulse characterisation, the output radiation is split so that 30\% is launched directly onto a 12~GHz photodetector (EOT ET5000f) to record time-domain evolution using a digital storage oscilloscope. The recorded 1D real-time intensity dynamics of the laser are then used to produce the 2D spatio-temporal intensity evolution, demonstrating both fast time dynamics over pulse roundtrip repetition time and slower propagation evolution over a number of cavity round trips. The same port has also been used for recording averaged spectra via an optical spectrum analyser (Yokogawa AQ6375B). The other 70\% of the generated signal has been analysed using DFT methodology in the spectral domain at the shot-to-shot level. The mapping of the temporal waveform of the generated pulses into the optical spectra has been performed using a pair of chirped fibre Bragg gratings (from Terraxion) connected in series by two fibre circulators. Due to the linear chirp of 24\,nm over their length of 140\,mm centred on 1905\,nm, the stretcher ensures a $\sim$100-ps\,nm$^{-1}$ group delay. This results in the elongating of the generated pulse from 0.8\,ps to 800\,ps, sufficient to satisfy the far-field condition and, therefore, for conversion in the spectral domain. 
Overall, the deployed measurement setup associated with a 33-GHz, 100-Gs/s digital storage oscilloscope (Tektronix DPO73304SX) and 22-GHz photodetector (from Discovery Semiconductors DSC2-30S) provides a 15-ps temporal resolution and a 0.1-nm spectral resolution, according to equations provided in Ref.~\cite{goda2013dispersive}.



\subsection{Numerical simulations}
To gain further insight into the formation dynamics of vector solitons in self-mode-locked Tm-doped fibre laser and oscillating dynamics of synchronised and unsynchronised dispersive waves, we have performed numerical simulations based on the system of coupled modified nonlinear Schr\"{o}dinger equations. The numerical model accounts for the effects of dispersion, birefringence, four-wave mixing and dynamically varying gain spectrum to describe the signal amplification:
\bigskip
\begin{gather}
\label{eq:prop}
\frac{\partial A_\mathrm{x}}{\partial z} =  i\frac{\Delta\beta}{2} A_\mathrm{x}- \delta \frac{\partial  A_\mathrm{x}}{\partial t} -i\frac{\beta_2}{2} \frac{\partial^2 A_\mathrm{x}}{\partial t^2} 
+ i\gamma \left( |A_\mathrm{x}|^2 + \frac{2}{3}|A_\mathrm{y}|^2\right) A_\mathrm{x} + \frac{1}{3} |A_\mathrm{y}|^2 A_\mathrm{x}^*
+ \int\limits_{-\infty}^\infty\frac{g_s(\omega,z)}{2}\tilde A_\mathrm{x}(z,\omega)e^{-i\omega t}d\omega,
 \\
\frac{\partial A_\mathrm{y}}{\partial z} = - i\frac{\Delta\beta}{2} A_\mathrm{y} + \delta \frac{\partial  A_\mathrm{y}}{\partial t} -i\frac{\beta_2}{2} \frac{\partial^2 A_\mathrm{y}}{\partial t^2} 
+ i\gamma \left( |A_\mathrm{y}|^2 + \frac{2}{3}|A_\mathrm{x}|^2\right) A_\mathrm{y} + \frac{1}{3} |A_\mathrm{x}|^2 A_\mathrm{y}^*
+ \int\limits_{-\infty}^\infty\frac{g_s(\omega,z)}{2}\tilde A_\mathrm{y}(z,\omega)e^{-i\omega t}d\omega, 
\\
\frac{\partial P_\mathrm{p}(z)}{\partial z} = g_\mathrm{p}(z)P_\mathrm{p}(z),
\end{gather}

\noindent where $A_\mathrm{x}$, $A_\mathrm{y}$ are the orthogonal components of the field envelope, $P_p(z)$ is the power of the continuous wave pump, $z$ is the longitudial coordinate, $t$ is the time. $\Delta \beta = 2\pi/L_B$ is the difference of the propagation constants due to fibre birefringence, and $L_B=L/40\approx12$ cm denotes the birefringence length. $\delta=\Delta \beta/2\omega_0$ is the inverse group velocity difference, $\omega_0$ is the central frequency. The central wavelength used in the simulations is 1800 nm. The Tm-doped gain fibre was simulated using following parameters: $\beta_2=-20.6$ ps$^2$/km is the group velocity dispersion, $\gamma = 2$ W$^{-1}$km$^{-1}$ is the Kerr nonlinearity, $g_s$ and $g_p$ are signal and pump gain/loss coefficients. The initial field for the first round trip was modelled as "white" Gaussian noise.

The frequency dependence of the gain $g_s(\omega, z)$ and pump gain/loss coefficient at each step along the fibre were found from the population inversion rate equations, as detailed in~\cite{kirsch2022gain}. A key distinction in the gain model employed in this work is that gain is saturated by the total power of both polarisation components of the field $P_s(z) = |A_x(z)|^2 + |A_y(z)|^2$.

The active fibre is considered as a combination of the amplifier described by the frequency-dependent gain profile $g_s(\omega, z)$ distributed over the whole fibre length and an absorber described as the time-dependent optical losses $\alpha(t,z)=-g_s(t,z)$ localised in a rear end of fibre section~\cite{kirsch2022gain}. The saturable absorption behaviour of the Tm-doped fibre segment was considered to have a modulation depth of 19\%, 81\% nonsaturable loss and saturation peak power of 118~W, as measured experimentally in our previous work~\cite{kirsch2022gain}.

For passive fibre, the parameters of group velocity dispersion and nonlinearity were used as $\beta_2=-59$ ps$^2$/km and $\gamma = 1.3$ W$^{-1}$km$^{-1}$, correspondingly.

\medskip

\medskip

\textbf{Acknowledgements}\par
D.C.K. and M.C. acknowledge the support of the Deutsche Forschungsgemeinschaft (DFG—German Research Foundation, Project No. CH 2600\_1-1). A.B. acknowledge the support of the Russian Science Foundation (24-12-00314, https://rscf.ru/en/project/24-12-00314/). 

\textbf{Conflict of Interest}\par 
The authors declare no conflict of interest.

\textbf{Data Availability Statement} \par 
The data that support the findings of this study are openly available in Figshare at \\http://doi.org/10.6084/m9.figshare.27951423, reference number 27951423.

\medskip


\bibliographystyle{MSP}
\bibliography{sample}







\end{document}